\documentclass[twocolumn]{aastex63}
\usepackage{graphicx}
\usepackage{multirow}

\shortauthors{Cai et al.}

\begin{document}

\title{Statistics of Solar White-Light Flares I: Optimization of Identification Methods and Application}

\author[0009-0007-4469-0663]{Yingjie Cai}
\affiliation{National Astronomical Observatories, Chinese Academy of Sciences, Beijing 100101, China}
\affiliation{School of Astronomy and Space Science, University of Chinese Academy of Sciences, Beijing 100049,
China}

\author[0000-0002-9534-1638]{Yijun Hou}
\affiliation{National Astronomical Observatories, Chinese Academy of Sciences, Beijing 100101, China}
\affiliation{School of Astronomy and Space Science, University of Chinese Academy of Sciences, Beijing 100049,
China}
\affiliation{Yunnan Key Laboratory of the Solar physics and Space Science, Kunming 650216, China}
\affiliation{State Key Laboratory of Solar Activity and Space Weather, Beijing 100190, China}

\author[0000-0001-6655-1743]{Ting Li}
\affiliation{National Astronomical Observatories, Chinese Academy of Sciences, Beijing 100101, China}
\affiliation{School of Astronomy and Space Science, University of Chinese Academy of Sciences, Beijing 100049,
China}
\affiliation{State Key Laboratory of Solar Activity and Space Weather, Beijing 100190, China}

\author{Jifeng Liu}
\affiliation{National Astronomical Observatories, Chinese Academy of Sciences, Beijing 100101, China}
\affiliation{School of Astronomy and Space Science, University of Chinese Academy of Sciences, Beijing 100049,
China}
\affiliation{Institute for Frontiers in Astronomy and Astrophysics, Beijing Normal University, Beijing 102206,
China}

\correspondingauthor{Yijun Hou}
\email{yijunhou@nao.cas.cn}

\begin{abstract}
White-light flares (WLFs) are energetic activity in stellar atmosphere. However, the observed solar WLF is relatively rare compared to
stellar WLFs or solar flares observed at other wavelengths, limiting our further understanding solar/stellar WLFs through statistical
studies. By analyzing flare observations from the \emph{Solar Dynamics Observatory (SDO)}, here we improve WLF identification methods
for obtaining more solar WLFs and their accurate light curves from two aspects: 1) imposing constraints defined by the typical temporal
and spatial distribution characteristics of WLF-induced signals; 2) setting the intrinsic threshold for each pixel in the flare ribbon
region according to its inherent background fluctuation rather than a fixed threshold for the whole region. Applying the
optimized method to 90 flares (30 C-class ones, 30 M-class ones, and 30 X-class ones) for a statistical study, we identified a total
of 9 C-class WLFs, 18 M-class WLFs, and 28 X-class WLFs. The WLF identification rate of C-class flares reported here reaches 30\%,
which is the highest to date to our best knowledge. It is also revealed that in each GOES energy level, the proportion of WLFs
is higher in confined flares than that in eruptive flares. Moreover, a power-law relation is found between the WLF energy (\emph{E})
and duration ($\tau$): $\tau \propto {E}^{0.22}$, similar to those of solar hard/soft X-ray flares and other stellar WLFs.
These results indicate that we could recognize more solar WLFs through optimizing the identification method, which will lay a base
for future statistical and comparison study of solar and stellar WLFs.
\end{abstract}

\keywords{Solar activity (1475); Solar atmosphere (1477); Solar flares (1496)}

\section{Introduction}\label{sect1}
Solar flares are powerful localized electromagnetic radiation bursts in the solar atmosphere, which emit radiation across the entire
electromagnetic spectrum \citep{1985srph.book...53M,2011LRSP....8....6S}. Although they are prominent in extreme
ultraviolet (EUV) and X-ray channels, a significant portion of solar flare energy is released at visible and UV wavelengths
\citep{2017LRSP...14....2B}. On 1859 September 1, the first solar flare ever observed in human history was seen by Richard Carrington
with the naked eye, which was later called as the Carrington flare \citep{1859MNRAS..20...13C,1859MNRAS..20...15H}. Such kind of
flares exhibiting a sudden enhancement of the continuum intensity at visible wavelength are later named white-light flares (WLFs)
\citep{1970SoPh...13..471S,1989SoPh..121..261N,1993SoPh..144..169N,2011SSRv..158....5H}. Compared with solar flares observed
in EUV and X-ray channels, WLFs usually have the following typical characteristics: they are quite rare in number
\citep{1966SSRv....5..388S,1983SoPh...88..275N}, possess relatively smaller spatial scales \citep{2018ApJ...867..159S}, and have
shorter durations \citep{2013RAA....13.1509F}.

In spite of more than 160 years having passed since the discovery of the first solar WLF, the exact generation mechanism of WLFs are
still under debate \citep{2016SoPh..291.1273H}. The current mainstream theories include soft X-ray (SXR)/EUV irradiation
\citep{1977A&A....57..105H,1988SoPh..115..277P}, proton beam bombardment
\citep{1978SoPh...58..363M,1993A&A...274..923H}, Alfv{\'e}n wave dissipation \citep{1982SoPh...80...99E,2008ApJ...675.1645F},
radiative backwarming \citep{1990ApJ...350..463M,2003A&A...403.1151D}, chromospheric condensation
\citep{1992ApJ...397..694G,2015SoPh..290.3487K}, nonthermal electron-beam bombardment
\citep{1992PASJ...44L..77H,2015ApJ...802...19K,2020ApJ...891...88W}. In addition, based on the spectral characteristics and the
relationship between WL emission and hard X-ray (HXR) or microwave emission, WLFs are generally categorized into two types
\citep{1986A&A...159...33M}, whose emission mechanisms and origins of WL emission are controversial
\citep{1999A&A...348L..29D}. The flare spectra of type \uppercase\expandafter{\romannumeral1} WLFs are characterized by a Balmer
jump and very strong emission of Balmer lines. Furthermore, the WL emission of type \uppercase\expandafter{\romannumeral1}
WLFs is almost synchronous with HXR or microwave emissions \citep{1995A&AS..110...99F}, which is believed to be produced by
hydrogen recombination in the chromosphere \citep{1984SoPh...92..217N}. On the other hand, type
\uppercase\expandafter{\romannumeral2} WLFs lack strong chromospheric Balmer continuum emissions and exhibit a time lag between
their WL emissions and HXR emissions \citep{1999ApJ...512..454D}. It is accepted that the WL emission of
type \uppercase\expandafter{\romannumeral2} WLFs is generated by the increased $\mathrm{H}^{\mathrm{-}}$ emissions in the
photosphere \citep{1982SoPh...80..113H,1985SoPh...98..255B}.

Although most previous studies indicated that WLFs are usually related to energetic solar eruptions like GOES X-class flares
\citep{2003A&A...409.1107M,2009RAA.....9..127W,2017ApJ...850..204W}, recent advances in high spatial resolution observations
have provided more opportunities to detect WLFs even in weak C-class flares
\citep{2018ApJ...867..159S,2018A&A...613A..69S,2020ApJ...904...96C}. It naturally raises a question that if the WL emission
enhancement is a common feature in all solar flares
\citep{1989SoPh..121..261N,2006SoPh..234...79H,2008ApJ...688L.119J,2018A&A...613A..69S}.
To answer this question, the key point is accurately identifying WLFs, especially for the weak ones. In most of past studies, the
WL emission enhancement signals were extracted by performing difference imaging technique on the WL continuum imaging
observations. For example, by amplifying the differences between two adjacent continuum images, a pseudo-intensity image is
constructed that can help us clearly identify the faint WL emission changes \citep{2018A&A...613A..69S}. \citet{2016ApJ...816....6K}
determined the location of WL emission enhancement signals by utilizing the spatial relationship between WL emission and HXR
emission generated by non-thermal electrons depositing energy through collisions in the lower layers of the solar atmosphere.

Based on the aforementioned identification methods and high-resolution WL observations from new generation solar telescopes like the
\emph{Solar Dynamics Observatory} \citep[\emph{SDO};][]{2012SoPh..275....3P}, some small-sample databases of solar WLFs were
established recently and statistical studies of WLFs have been thus conducted
\citep{2015SoPh..290.3151B,2016ApJ...816....6K,2016RAA....16..177H,2017ApJ...850..204W,2018A&A...613A..69S,2018ApJ...867..159S,2020ApJ...904...96C}.
It was statistically revealed that there is a certain degree of correlation between WL emission and changes in the
photospheric magnetic field during flares \citep{2020ApJ...904...96C}. Moreover, \citet{2018ApJ...867..159S} performed the first
statistical investigation of WL emission in circular-ribbon flares and found that these circular-ribbon WLFs generally exhibit shorter
durations, smaller spatial extents, intensified electric currents, and intricate magnetic field configurations.

Although more and more solar WLFs have been discovered and investigated in recent statistical studies, to our best knowledge, the
largest WLF database in one single study contains no more than one hundred WLFs. Such fact stands in stark contrast to the situation
on other stars. When we turn our gaze to other distant stars through the Kepler observations, a considerable amount of stars are found
to exhibit numerous WLFs during only a few years, which is equivalent to (if not more than) the number of WLFs detected in the Sun
over the past hundred years \citep{2012Natur.485..478M,2015EP&S...67...59M,2021ApJ...922L..23A,2021ApJ...906...72O}.
Therefore, in recent years, extensive efforts have been made in comparative studies between these stellar flares and solar flares. It has
been widely accepted that solar and stellar flares have similar observational characteristics and they might share the same mechanism
of energy release, i.e., magnetic reconnection \citep{2015EP&S...67...59M,2021MNRAS.505L..79Y,2024LRSP...21....1K}.
For example, \citet{2017ApJ...851...91N} carried out a statistical study on 50 solar WLFs and found that the correlation between
the energies ($E$) and durations ($\tau$) of these solar WLFs ($\tau \propto {E}^{0.38}$) is similar to that on stellar superflares.
Even more impressive is their attempt to diagnose the physical parameters like magnetic field strength of flare core region through the
observed $E$-$\tau$ diagram of flares. However, the credible power-law relation between solar WLF energy and duration should be derived
from a large enough sample of solar WLFs, which is not available there. As a result, it is urgent to construct a large sample of solar WLFs
and then obtain their accurate parameters like energy and duration for further solar WLF studies.

One can see that the limited number of detected solar WLFs has posed a significant obstacle to in-depth investigations into the
mechanisms behind their formation, as well as to conducting further statistical comparisons between solar and stellar WLFs. According
to the previous finding of weak C-class WLFs \citep{1989SoPh..121..261N,2006SoPh..234...79H,2008ApJ...688L.119J,2018A&A...613A..69S,
2020ApJ...904...96C}, we tend to believe that WLFs are not a rare phenomenon in the Sun, and the previous scarcity in the number of
solar WLFs could be attributed to limitations in observational data resolution or flaws in identification methods. As a result,
in the present study, we aim to improve the solar WLF identification method through conducting an in-depth analysis of optical continuum
observations during flares from the \emph{SDO}, by far the best telescope for solar flare research. Based on such optimized methods,
we would have the ability to identify more solar WLFs and obtain their more accurate integral WL intensity light curves, from which solar
WLF physical parameters like energy and duration would be derived. This will lay the foundation for establishing a large database of
solar WLFs and subsequently conducting direct comparative statistical studies between solar and stellar WLFs.

The remainder of this paper is structured as follows. We describe the employed observations and analysis in Section \ref{sect2}. The
typical spatial and temporal distributions of WL emission enhancement signals produced by WLFs and the according optimization of WLF
identification method are presented in Section \ref{sect3}. In Section \ref{sect4}, we identify 55 WLFs among 90 solar flares
and conduct a statistical analysis of these obtained solar WLFs. Finally, we give a brief summary in Section \ref{sect5}.

\section{Observations and Data Analysis}\label{sect2}
Our primary data sources are the Helioseismic and Magnetic Imager \citep[HMI;][]{2012SoPh..275..207S} and the Atmospheric
Imaging Assembly \citep[AIA;][]{2012SoPh..275...17L} aboard \emph{SDO}. We used the hmi.Ic$\_45s$ (continuum intensity) with a
cadence of 45 s and plate scale of $0.504^{\prime \prime}$ pixel\textsuperscript{-1}. The HMI continuum data refers to map of the
continuum intensity of the solar spectrum around Fe \uppercase\expandafter{\romannumeral1} absorption line at $6173\,\text{\AA}$ on
the surface of the Sun. The AIA was constructed to capture comprehensive images of the solar atmosphere, ensuring a field of view of
no less than $40^{\prime}$ and a spatial pixel resolution of $0.6^{\prime \prime}$. The telescopes are equipped with filters that
encompass ten distinct wavelength bands. These bands include seven extreme ultraviolet, two ultraviolet, and one visible wavelengths.
The time cadences for the EUV and UV observations are 12 s and 24 s. Here we only used $1600\,\text{\AA}$ images with a
cadence of 24 s to approximately determine the flare ribbon region, where WL emission enhancement signals might appear.

In addition, we also constructed a solar WLF sample by applying the improved WLF identification methods proposed here to 90
flares (30 C-, 30 M-, and 30 X-class flares) recorded by GOES SXR observations. Among these 90 flares, 24 X-, 30 M-, and 30 C-class
ones were randomly selected from 719 flares (above C5.0) studied in \citet{2021ApJ...917L..29L}, where the coronal mass ejection (CME) association
for the flare was determined. Another 6 X-class flares were selected from recent observations since 2021, whose CME associations
were determined based on the observations from the Large Angle and Spectrometric Coronagraph \citep[LASCO;][]{1995SoPh..162..357B}
on board the Solar and Heliospheric Observatory satellite \citep[SOHO;][]{1995SoPh..162....1D}. From these 90 flares, we identified
55 WLFs and carried out statistical analyses on them.

The continuum images of all the investigated flares have been corrected for limb darkening to second order \citep{2020ApJ...904...96C}.
The corrected intensity of each pixel is given by
\begin{equation}
\ I_{ij}^{corr} = \frac{I_{ij}^{non - corr}}{C_{ij}}, \label{eq:1}
\end{equation}
where $C_{ij}$ refers to the limb darkening function
\begin{equation}
\ C_{ij}=1-u_\lambda-\nu_\lambda+u_\lambda\cos(\Theta)+\nu_\lambda\cos(\Theta)^2,  \label{eq:2}
\end{equation}
where $\Theta = \sin^{-1}(\sqrt{(x_{i}-x_{c})^{2}+(y_{j}-y_{c})^{2}}/R_{\odot})$. $(x_{i},y_{i})$ and $(x_{c},y_{c})$ respectively
refer to the given pixel coordinates and coordinates of the solar disk center. For the hmi.Ic$\_45s$ (continuum intensity), the
wavelength-dependent parameters $u_\lambda$ and $\nu_\lambda$ are respectively equal to $u_{6173.3}=0.836$ and $\nu_{6173.3}=-0.204$
\citep{1976asqu.book.....A}. Furthermore, it is worth noting that for each flare, we removed the effects of solar rotation by aligning
their continuum images to a common time. This approach helps us avoid the risk of false WL emission enhancement signals arising from
shifts of features caused by solar rotation.

\section{Optimization of WLF identification methods}\label{sect3}
As mentioned in the Introduction, traditional WLF identification methods are essentially based on the difference imaging technique,
which can highlight WL emission enhancement in WLFs. Then a fixed threshold of the difference contrast needs to be set for
screening WL emission enhancements. However, such traditional methods have two inevitable flaws: 1) if the threshold is set
too high, weak WL emission enhancement produced by less energetic flares (e.g., C-class flares) would be missed; 2) if the threshold is
set too low,  background WL emission enhancement with inherent fluctuation caused by constant convective motion would be
significantly introduced. As a result, we aim to develop new WLF identification methods for solar flares with different GOES energy
levels, which can accurately identify the weak WL emission enhancement while avoid involving the background WL emission fluctuation
signals.

\subsection{Typical spatial and temporal distribution of WL enhancement signals produced by WLFs}\label{sect31}

To propose optimized WLF identification methods, the first thing we need to do is understanding the typical spatial and temporal
distribution of WL emission enhancement in WLFs. Here three solar flares confirmed as WLFs in previous studies were selected to
explore if there are any common characteristics among WLFs. In order to ensure the universality of the results, the three WLFs have
different GOES energy levels: one C8.6 WLF \citep{2018ApJ...867..159S}, one M6.6 WLF \citep{2020ApJ...904...96C}, and one X9.3
WLF \citep{2017ApJ...849L..21Y,2018A&A...619A.100H,2018ApJ...856...79Y}. For each WLF, we firstly determined the flare ribbon
region according to the AIA $1600\,\text{\AA}$ observations. Then we made running difference images based on the HMI continuum
intensity observations during half an hour around the GOES peak time of the flare. Finally, a fixed threshold was set, and pixels with
qualified WL emission enhancement at each difference image were recorded.

\begin{figure*}[htbp]
\centering
\includegraphics [width=1\textwidth]{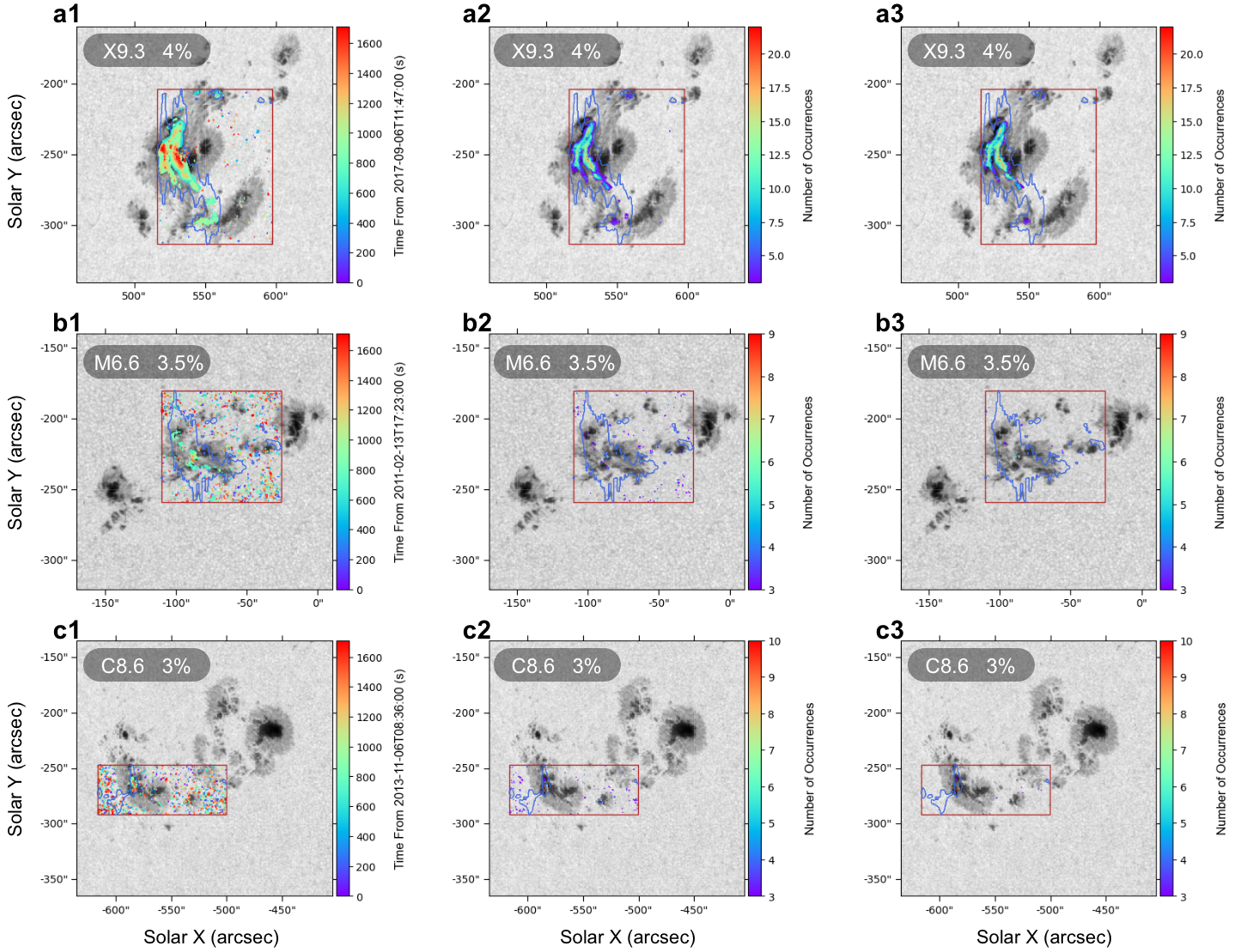}
\caption{Spatial and temporal distributions of WL emission enhancement signals produced by three WLFs with different GOES energy
levels. (a1): Spatial distribution of the pixels with WL emission enhancement larger than 4\% during the X9.3 WLF. The appearance time
of these enhancement signals is marked by different colors. The blue contour approximates the flare ribbon region based on the
AIA $1600\,\text{\AA}$ images with a cadence of 24 s and the red rectangle encompasses the surrounding quiet flare region of the flare
ribbon region. (a2): Spatial distribution of the pixels where WL emission enhancement larger than 4\% appears three times or more in
total during the X9.3 WLF. The number of occurrences is marked by different colors.  (a3): Spatial distribution of the pixels where WL
emission enhancement larger than 4\% successively appears three times or more during the X9.3 WLF. The number of occurrences is
marked by different colors. (b1)-(c3): Similar to (a1)-(a3), but for the M6.6 and C8.6 WLFs, respectively.}
\label{fig1}
\end{figure*}

Figure \ref{fig1} presents spatial and temporal distributions of WL emission enhancement produced by the three WLFs with different
GOES energy levels. Top three panels in Figure \ref{fig1} show results of the X9.3 WLF. In Figure \ref{fig1}(a1), we mark all pixels with
WL emission enhancement larger than 4\% during the X9.3 WLF by using different colors to represent their appearance time. Compared
with previous studies \citep{2009RAA.....9..127W}, here we set a lower threshold of the difference contrast for ensuring there are
enough qualified pixels both in the flare ribbon region and quiet Sun region, which could make it possible to explore the distinction
between the WL emission enhancement signals produced by WLF and the background WL emission enhancement signals caused by
constant convective motion. In Figure \ref{fig1}(a2), pixels where WL emission enhancement larger than 4\% appears three times or
more in total during the X9.3 flare are marked with different colors to represent the number of their occurrences. In Figure \ref{fig1}(a3),
pixels where WL emission enhancement larger than 4\% appears successively three times or more during the X9.3 flare are marked
with different colors to represent the number of their occurrences. The middle and bottom rows in Figure \ref{fig1} are similar to
Figure \ref{fig1}(a1)--Figure \ref{fig1}(a3), but for the M6.6 and C8.6 WLFs. Considering different GOES energy levels of the three
WLFs, the fixed thresholds for the latter two WLFs are set as 3.5\% and 3\%, respectively.

From Figure \ref{fig1}(a1), Figure \ref{fig1}(b1), and Figure \ref{fig1}(c1), we can see that there are a considerable number of pixels
with WL emission enhancement meeting the threshold criteria in both the flare ribbon region and quiet Sun region. However, compared to
the signals in the quiet Sun region, those in the flare ribbon region exhibit an evident feature of spatial cluster, i.e., WL emission
enhancement signals appearing in different times are concentrated in some small regions (also shown in \citet{2020ApJ...904...96C}).
The middle and right columns in Figure \ref{fig1} reveal a substantial reduction of enhancement signals in both the quiet Sun region and
the flare ribbon region while the enhancement signals in the flare ribbon region still exhibit significant spatial clustering. Therefore,
we can conclude that during solar WLFs with different GOES energy levels, the WL emission enhancement signals produced by WLFs exhibit
pronounced spatial clustering characteristic and tend to occur repeatedly or consecutively in the time domain. But for the background WL
emission enhancement signals caused by the constant convective motion in the quiet Sun region, the constraint from the time domain would
significantly reduce their number of appearance. Based on these spatial and temporal distribution characteristics, we could try to improve
the WLF identification method for accurately detecting the weak WL emission enhancement signals produced by WLF while avoid involving the
background WL emission enhancements.

\begin{figure*}[htbp]
\centering
\includegraphics [width=1\textwidth]{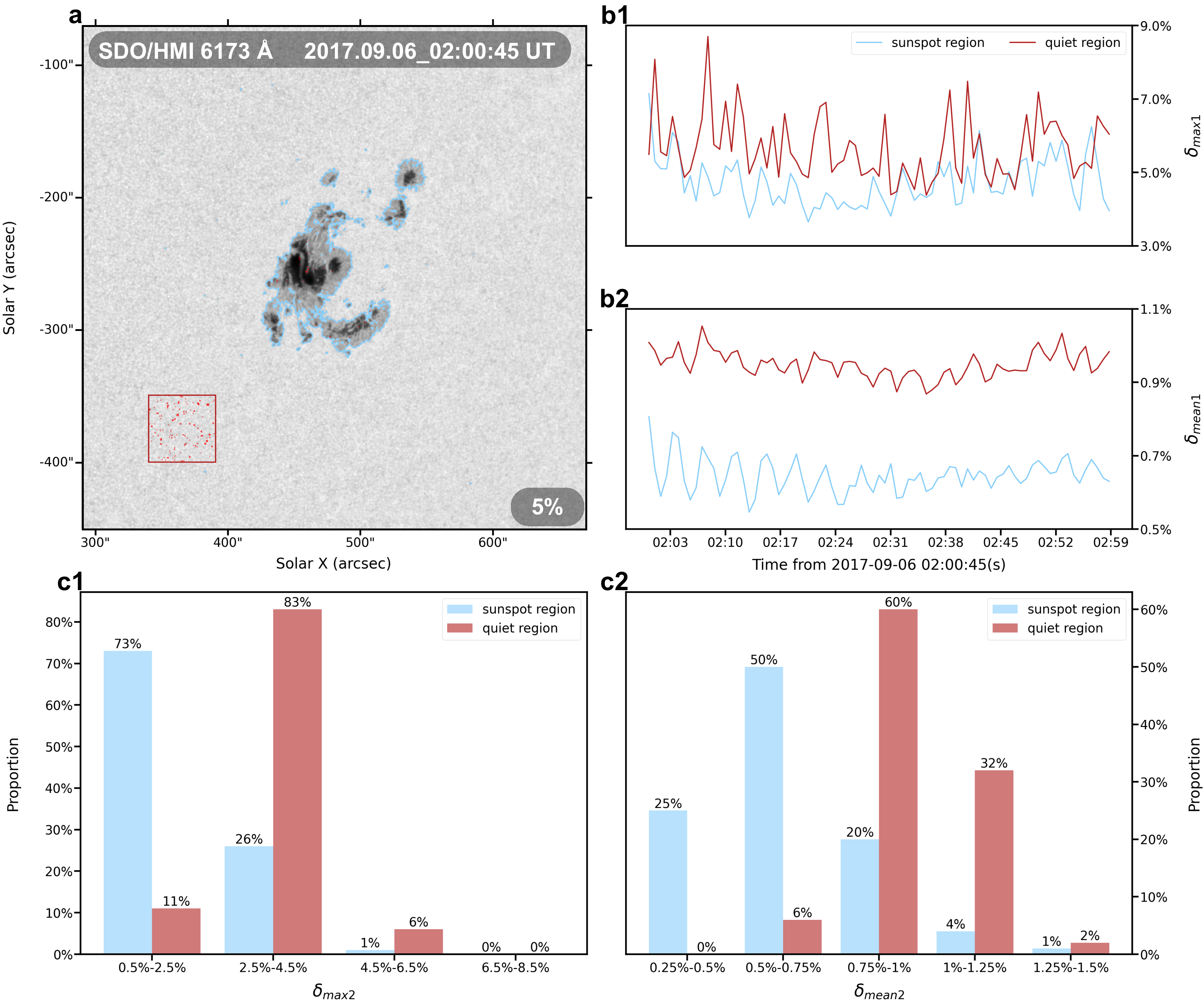}
\caption{Background WL emission with inherent fluctuation caused by constant convective motion in different regions around the
X9.3 flare ribbon region. (a): Spatial distribution of the pixels with WL emission enhancement larger than 5\% during one hour with no
occurrence of flare (2017-09-06 02:00 UT to 03:00 UT) in two different regions. The red rectangle and blue contour denote the quiet Sun
region and sunspot region, respectively. (b1): Temporal variation of the maximum fluctuation value among all the pixels in quiet Sun
region and sunspot region at each moment ($\delta_{max1}$). (b2): Similar to (b1), but for the mean fluctuation value of all the pixels in
each region ($\delta_{mean1}$). (c1): The proportion histogram of the maximum fluctuation value of every pixel during one hour in quiet
Sun region and sunspot region ($\delta_{max2}$). (c2): Similar to (c1), but for the mean fluctuation value of the pixels during one
hour ($\delta_{mean2}$).}
\label{fig2}
\end{figure*}

\begin{figure*}[htbp]
\centering
\includegraphics [width=1\textwidth]{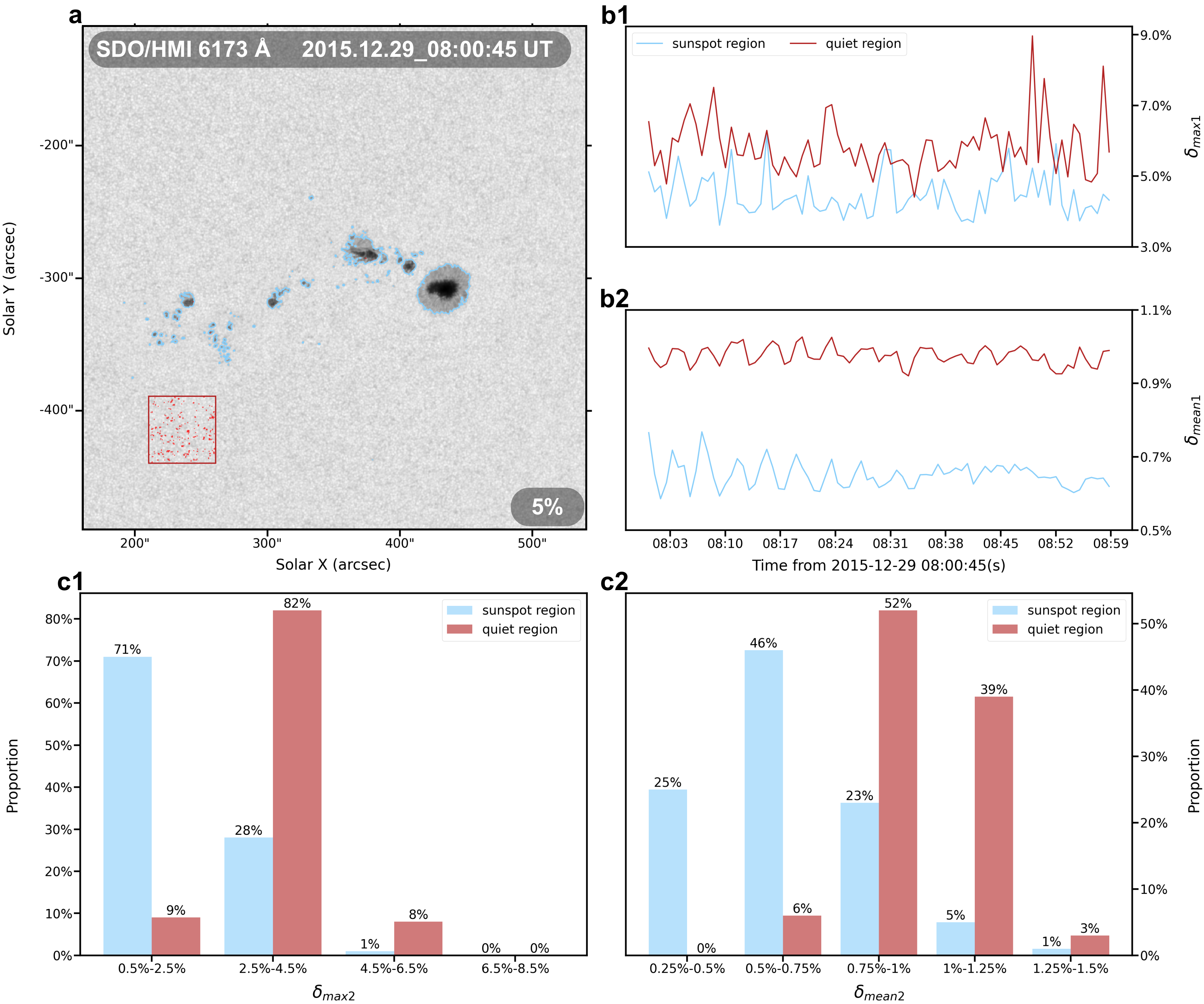}
\caption{Background WL emission with inherent fluctuation caused by constant convective motion in different regions around the
C6.8 flare ribbon region. (a): Spatial distribution of the pixels with WL enhancement larger than 5\% during one hour with no occurrence
of flare (2015-12-29 08:00 UT to 09:00 UT) in two different regions. The red rectangle and blue contour denote the quiet Sun region and
sunspot region, respectively. (b1): Temporal variation of the maximum fluctuation value among all the pixels in quiet Sun region and
sunspot region at each moment ($\delta_{max1}$). (b2): Similar to (b1), but for the mean fluctuation value of all the pixels in each
region ($\delta_{mean1}$). (c1): The proportion histogram of the maximum fluctuation value of every pixel during one hour in quiet
Sun region and sunspot region ($\delta_{max2}$). (c2): Similar to (c1), but for the mean fluctuation value of the pixels during one
hour ($\delta_{mean2}$).}
\label{fig3}
\end{figure*}

\subsection{Setting the intrinsic threshold}\label{sect32}
As mentioned at the beginning of Section \ref{sect3}, if a fixed threshold of difference contrast is set for identifying WLF, it will be
difficult to identify the real WLF signals from numerous false background signals. Moreover, the calculation of WLF light curve would
also be severely impacted by the introduction of background signals. On one hand, since the background WL emission enhancements
vary across different flares, it is unreasonable to apply the same threshold to different flare events. On the other hand, even in the
same flare, the background WL emission fluctuations in regions with different magnetic field and plasma conditions are distinct.
As a result, here we proposed a concept of intrinsic threshold for each pixel defined according to background WL emission with inherent
fluctuation caused by constant convective motion to further improve the WLF identification method. A similar processing method was also
once employed by \citet{2020ApJ...904...96C}. To quantitatively define the intrinsic threshold, we analyzed the background WL emission
with inherent fluctuation in different regions during one hour before the onset of X9.3 and C8.6 flares in Figures \ref{fig2} and \ref{fig3},
respectively.

The WL emission enhancement ($\delta_n$) for each pixel at every moment was firstly calculated according to the following formula:
\begin{equation}
\delta_{n} = \left|\frac{I_{n+1}-I_{n}}{I_{n}}\right|, \label{eq:3}
\end{equation}
where $I_{n}$ represents the WL continuum intensity value of a certain pixel at a specific moment, and $n$ indicates the current frame
of the image. As shown in Figure \ref{fig2}(a), when the traditional method with a fixed threshold of 5\% \citep{2018A&A...613A..69S} is
applied to the SDO/HMI observations during one hour before the onset of X9.3 flare, there are much less qualified WL emission
enhancement signals produced by the background fluctuation in sunspot region than those in surrounding quiet Sun region. It intuitively
indicates the inherent fluctuation of background WL emission in different regions is different with each other. To further quantify such
difference, in Figure \ref{fig2}(b1), we firstly analyzed the temporal variation of the maximum fluctuation value among all the pixels in
quiet Sun region and sunspot region at each moment. According to Equation \ref{eq:3}, $\delta_n$ of each pixel can be calculated in
the quiet Sun region and sunspot region. For each moment, the maximum value of $\delta_n$ among all the pixels in each region is
then marked as $\delta_{max1}$. Additionally, we also calculated the average value ($\delta_{mean1}$) of $\delta_n$ of all the pixels
in each region at each moment and obtained their temporal variations (see Figure \ref{fig2}(b2)). It is obvious that for both of the
$\delta_{max1}$ and $\delta_{mean1}$, the value of the quiet Sun region is generally larger than that of the sunspot region. This
implies that the background fluctuation in the quiet Sun region is stronger than that in the sunspot region which could be due to the
strong inhibition of convection in the sunspot region. Furthermore, in Figure \ref{fig2}(b2), we can find the p-modes solely by eyes
\citep{1962ApJ...135..474L,1992ApJ...387..372G,1994ApJ...424..466G,2001ApJ...555..990B}. In Figure \ref{fig2}(c1), 
we calculated $\delta_n$ of each pixel in the quiet Sun region and sunspot region between every two consecutive images during one hour 
before the onset of X9.3 flare, and the maximum value of the pixel is then marked as its $\delta_{max2}$. In addition, the average 
value of the pixel's $\delta_n$ during one hour before the onset of X9.3 flare is marked as $\delta_{mean2}$ (see Figure \ref{fig2}(c2)). 
It is obvious that for both of the $\delta_{max2}$ and $\delta_{mean2}$, the values of pixels in the quiet Sun region are generally 
larger than those of the sunspot region, which also implies that the pixels of the quiet Sun region exhibit stronger background fluctuations.

Furthermore, Figure \ref{fig3} exhibits results during one hour before the onset of C6.8 flare that are remarkably similar to Figure
\ref{fig2}. Based on Figures \ref{fig2} and \ref{fig3}, we can conclude that the background fluctuation in the sunspot region is much
weaker than that in the quiet Sun region. Therefore, a fixed threshold higher than the background fluctuation in the quiet Sun region
as employed in the traditional method will inevitably result in the following drawbacks: 1) WLF-related signals weaker than the
background fluctuation of quiet Sun region will be impossible to be identified; 2) occasional strong WL emission enhancements
produced by the constant convection motion will be incorrectly introduced. For example, the formation or dynamical evolution of
sunspot light bridges can usually cause intermittent WL emission enhancements within the sunspot
\citep{2018ApJ...854...92T,2020A&A...642A..44H,2022ApJ...929...12H}, which can be detected as real signals during the flares but are
actually not associated with the flares.

The aforementioned results request us to accordingly set an intrinsic threshold for each pixel in the improved WLF identification
methods. We replaced the fixed threshold for improved WLF identification methods in Section \hyperref[sect32]{3.2} by the intrinsic
threshold for each pixel according to the value of $\delta_{max2}$ during its quiet period. After repeated tests, we set the intrinsic
thresholds of X-class, M-class, and C-class flares as $2 \times \delta_{max2}$ or $1.5 \times \delta_{max2}$.

\subsection{Optimizations based on the typical characteristics and intrinsic threshold}\label{sect33}
According to the aforementioned results in Sections \hyperref[sect31]{3.1} and \hyperref[sect32]{3.2}, we proposed new WLF
identification methods as follows: 1) imposing constraints defined by the typical temporal and spatial distribution characteristics of WLF-
related signals; 2) setting the intrinsic threshold for each pixel in the flare region according to its inherent background fluctuation.
The methods we proposed are all based on running difference between two consecutive WL intensity images. Pixels with
$\delta_{n}$ larger  than the set threshold are identified to have WL emission enhancement signals and recorded. Then five different
methods were defined to identify the WL emission enhancement signals during the WLF, and their specific descriptions are listed as
follows:

\begin{enumerate}
\item Method 1 is the unimproved traditional approach with a fixed threshold used for comparison with the optimized WLF identification methods proposed here. The pixel with $\delta_n$ larger the fixed threshold once during the whole flare would be marked as a valid WL emission enhancement signal.
\item In Method 2, only the pixel that exhibits $\delta_n$ larger than the intrinsic threshold three times or more during the flare would be identified as a valid WL emission enhancement signal produced by the WLF. It is because that the WLF-related signals tend to occur repeatedly in the time domain.
\item In Method 3, only the pixel that exhibits $\delta_n$ larger than the intrinsic threshold continuously three times or more during the flare would be identified as a valid WL emission enhancement signal produced by the WLF. The continuous occurrence of valid WL emission enhancement signals produced by the WLF in the time domain is the reason for this method.
\item Considering the fact that the lifetime of WLF-related emission enhancement signals in some weak WLFs could be shorter than the cadence of HMI observation (45 s), Methods 2 and 3 will be unable to identify such signals. As a result, in Method 4, we try to identify WLF-related signals only based on their spatial clustering characteristics without temporal restrictions. Specifically, a pixel will be identified as a valid WL emission enhancement signal when at least five of its nine associated pixels (the central pixel and its surrounding eight pixels) exhibit $\delta_n$ larger than the intrinsic threshold at a given moment.
\item As a supplementary to Method 4, in Method 5, we consider two consecutive moments and identify a pixel to be a valid WL emission enhancement signal when at least nine of its eighteen associated pixels (the central pixel and its surrounding pixels at two moments) exhibit $\delta_n$ larger than the intrinsic threshold.
\end{enumerate}

After imposing the restrictions from the temporal or spatial distribution characteristics of WLF-related emission enhancement signals
and considering the intrinsic threshold, the aforementioned optimized methods can set a relatively small threshold and thus identify faint
WL emission enhancement signals produced by weak WLFs while minimizing the influence of background WL emission enhancement
caused by constant convective motion.

\begin{figure*}[htbp]
\centering
\includegraphics [width=1\textwidth]{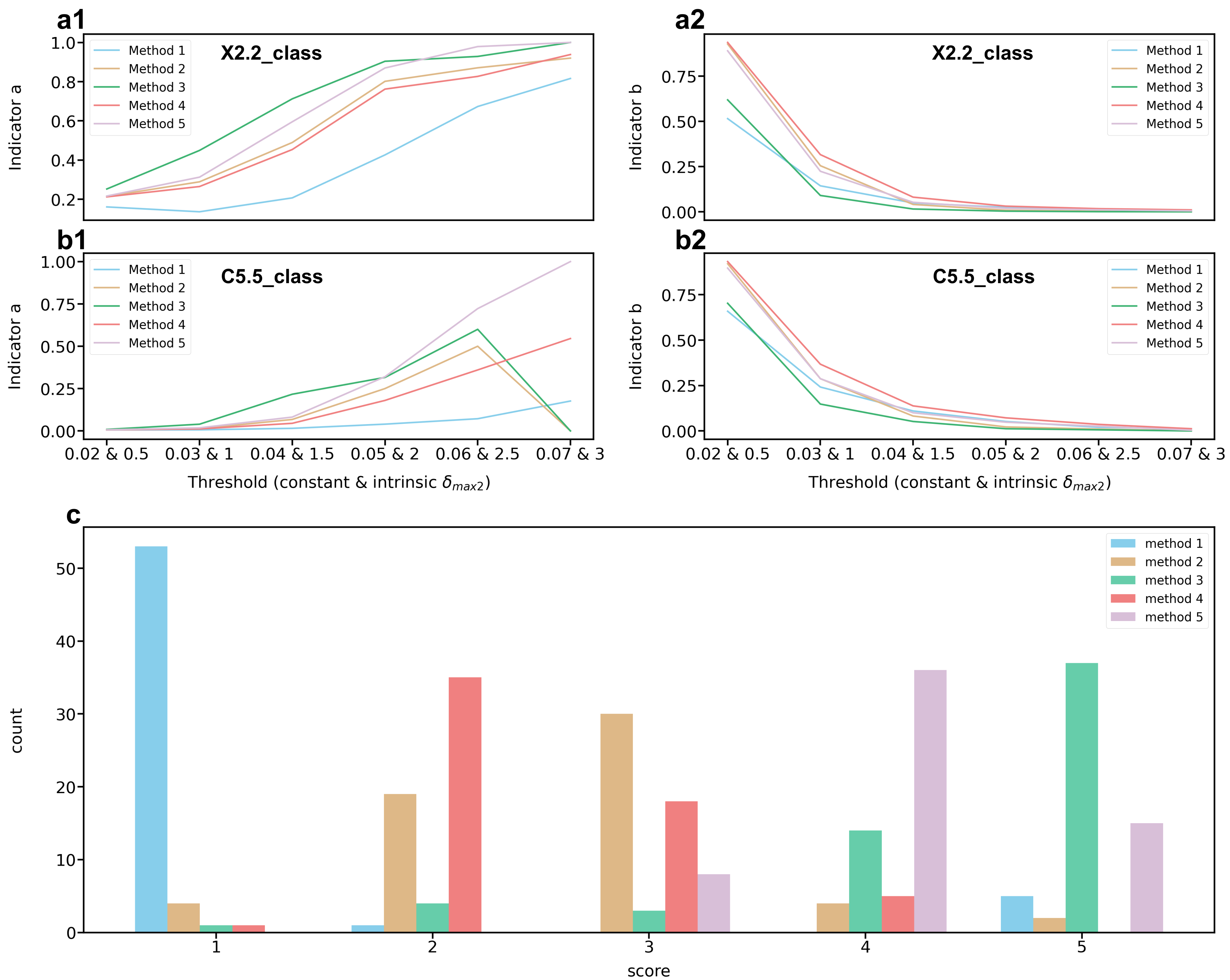}
\caption{Quantitative comparison of the five methods by consistent metrics. (a1): The variation of \emph{Indicator a} obtained by
the five methods under different thresholds for a X2.2-class WLF. Except for Method 1, all the other methods use intrinsic threshold
$\delta_{max2}$. (a2): The variation of \emph{Indicator b} obtained by the five methods under different thresholds for a X2.2-class WLF.
(b1)-(b2): Similar to (a1)-(a2), but for a C5.5-class WLF. The left side of the horizontal axis ticker represents the fixed thresholds used by
Method 1, while the right side represents the intrinsic threshold $\delta_{max2}$ used by other methods.  (c): Quantitative assessment
of \emph{Indicator a} for the five methods based on 10 WLFs.}
\label{fig4}
\end{figure*}

\begin{figure*}[htbp]
\centering
\includegraphics [width=1\textwidth]{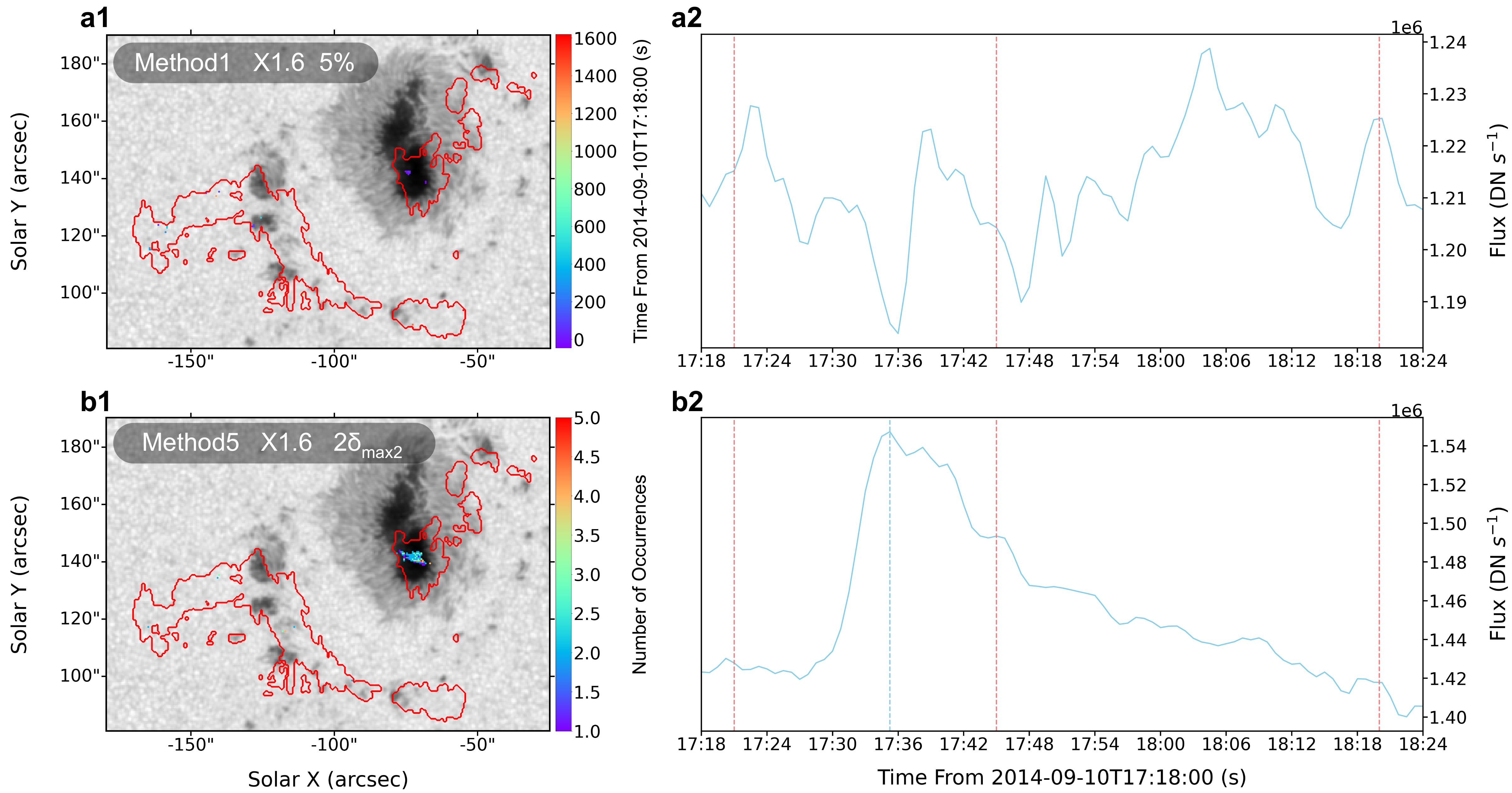}
\caption{Performance comparison between methods 1 and 5 when applied to an X1.6-class WLF. (a1): Spatial distribution of
the pixels with WL emission enhancement identified by Method 1 (with a constant threshold of 5\%) during the X1.6-class WLF. The
appearance time of the enhancement is marked by different colors. The red contour encompasses the flare ribbon region based on
the AIA $1600\,\text{\AA}$ images. (a2): Calculated WL light curve profile during the flare. The red vertical lines in panels (a2) and
(b2) denote the start, peak, and end times of GOES SXR $1-8\,\text{\AA}$. (b1)-(b2): Similar to (a1)-(a2), but for Method 5 (with a intrinsic
threshold of $2 \times \delta_{max2}$). The blue vertical line in panel (b2) denotes the peak time of calculated WL light curve.}
\label{fig5}
\end{figure*}

\begin{figure*}[htbp]
\centering
\includegraphics [width=1\textwidth]{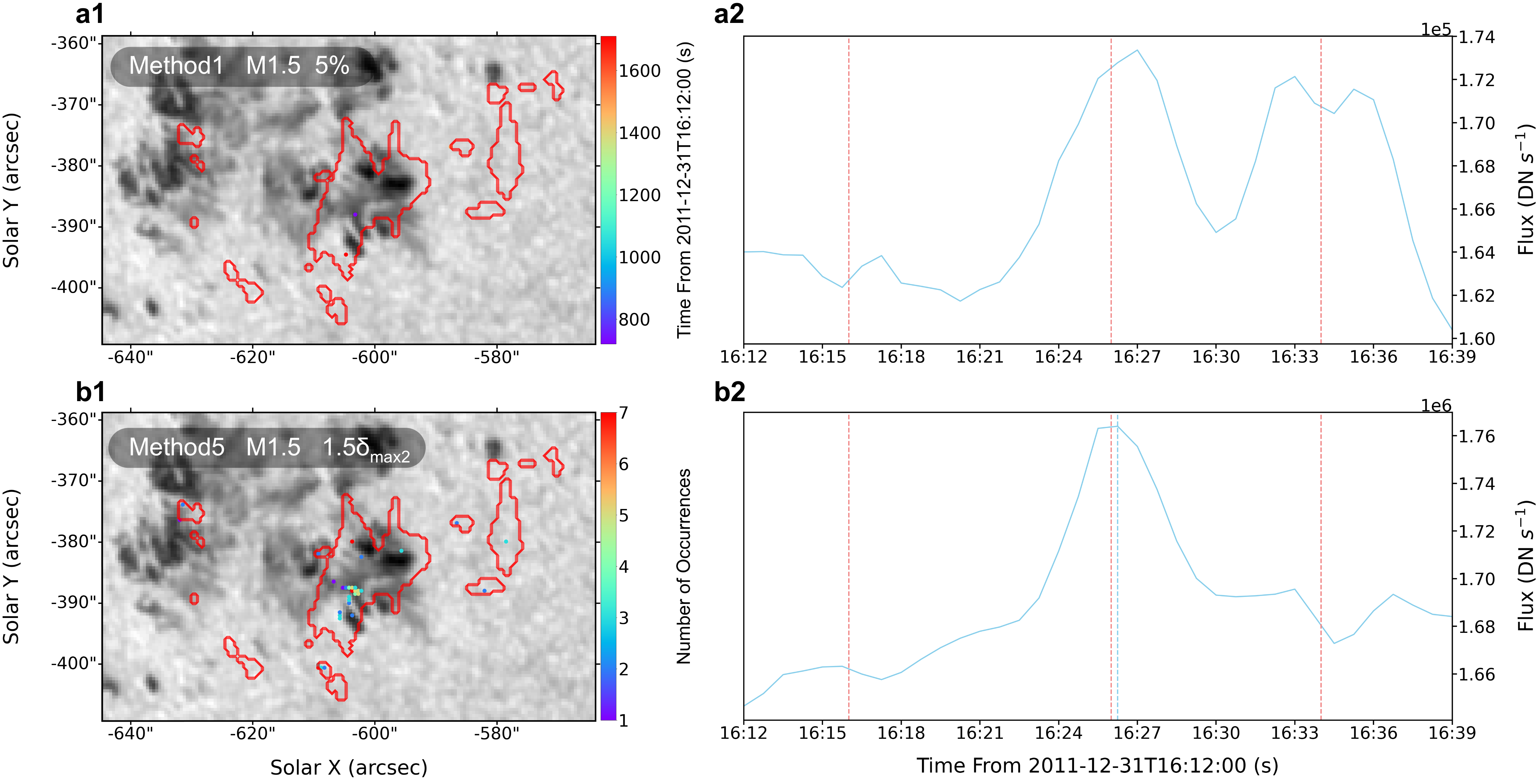}
\caption{Performance comparison between methods 1 and 5 when applied to a M1.5-class WLF. (a1): Spatial distribution of
the pixels with WL emission enhancement identified by Method 1 (with a constant threshold of 5\%) during the M1.5-class WLF. The
appearance time of the enhancement is marked by different colors. The red contour encompasses the flare ribbon region based on
the AIA $1600\,\text{\AA}$ images. (a2): Calculated WL light curve profile during the flare. The red vertical lines in panels (a2) and
(b2) denote the start, peak, and end times of GOES SXR $1-8\,\text{\AA}$. (b1)-(b2): Similar to (a1)-(a2), but for Method 5 (with a
intrinsic threshold of $1.5 \times \delta_{max2}$). The blue vertical line in panel (b2) denotes the peak time of calculated WL light
curve.}
\label{fig6}
\end{figure*}

\begin{figure*}[htbp]
\centering
\includegraphics [width=1\textwidth]{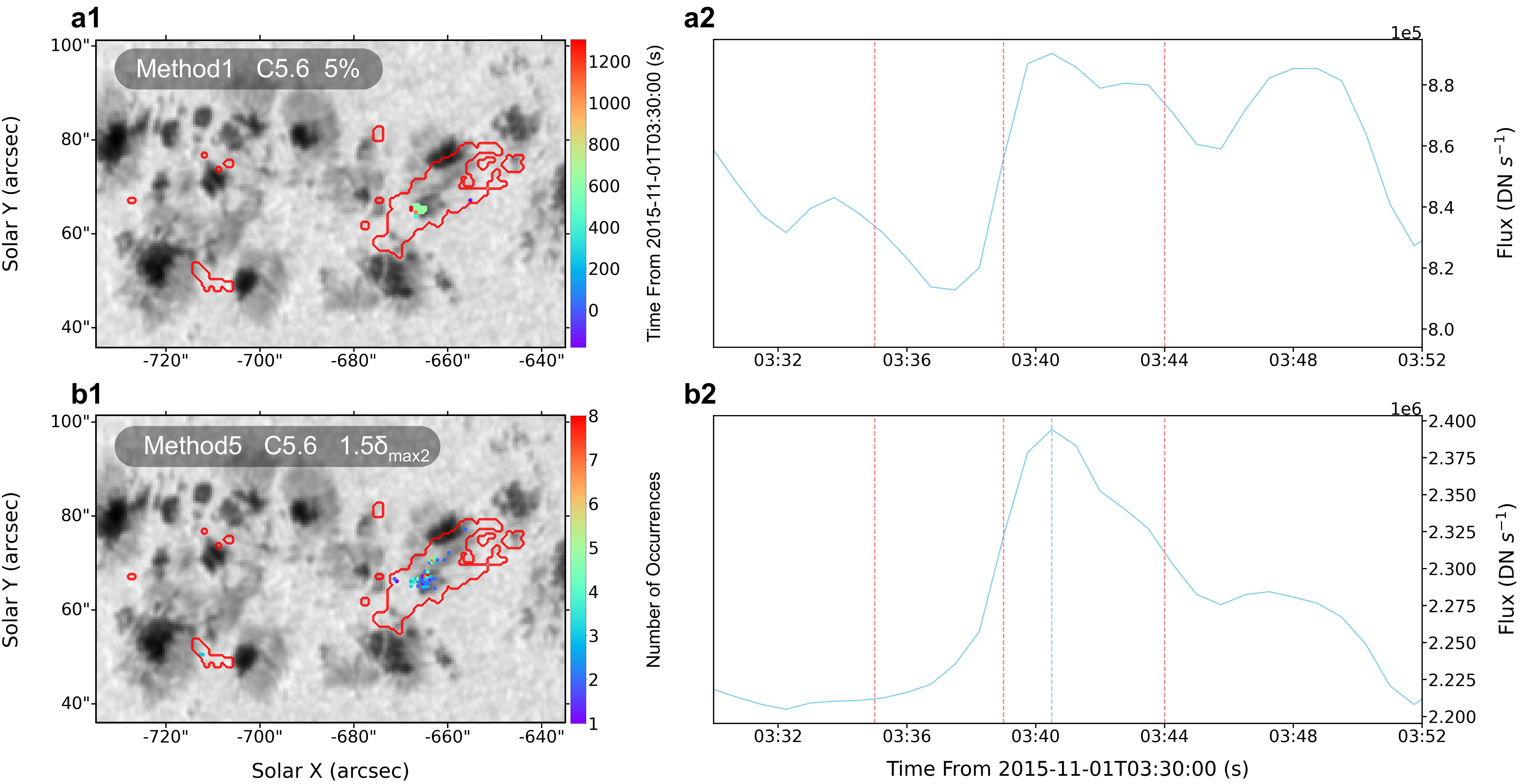}
\caption{Performance comparison between methods 1 and 5 when applied to a C5.6-class WLF. (a1): Spatial distribution of
the pixels with WL emission enhancement identified by Method 1 (with a constant threshold of 5\%) during the C5.6-class WLF. The
appearance time of the enhancement is marked by different colors. The red contour encompasses the flare ribbon region based on
the AIA $1600\,\text{\AA}$ images. (a2): Calculated WL light curve profile during the flare. The red vertical lines in panels (a2) and
(b2) denote the start, peak, and end times of GOES SXR $1-8\,\text{\AA}$. (b1)-(b2): Similar to (a1)-(a2), but for Method 5 (with a
intrinsic threshold of $1.5 \times \delta_{max2}$). The blue vertical line in panel (b2) denotes the peak time of calculated WL light
curve.}
\label{fig7}
\end{figure*}

\begin{figure*}[htbp]
\centering
\includegraphics [width=1\textwidth]{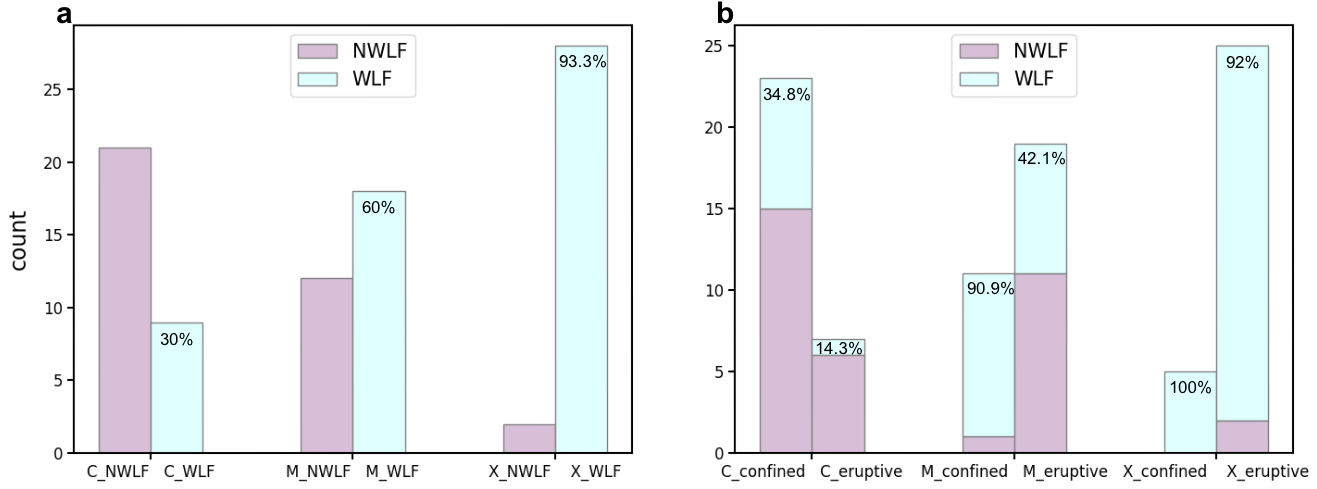}
\caption{Percentage of WLFs among different GOES energy levels and two flare types (eruptive and confined). (a): Percentage
of WLFs among C-class, M-class, and X-class flares, respectively. The purple histograms mark the flares that are not WLFs (NWLF),
and the cyan histograms mark the WLFs. (b): Percentage of WLFs among confined and eruptive C-class, M-class, and X-class flares,
respectively.}
\label{fig8}
\end{figure*}

\subsection{Comparison between different methods}\label{sect34}
In order to compare the effectiveness of different methods proposed here and obtain a consistent metric to determine the best
method, we chose 10 potential WLFs from our flare catalog based on visual identification for quantitative analysis. According
to the definition of WL emission enhancement signal in different methods, we firstly obtained qualified pixels during each WLF for each
method. As shown in Figure \ref{fig1}, based on AIA $1600\,\text{\AA}$ observations, we then marked flare ribbon regions by blue
contours and the surrounding regions with red rectangles. Here we have an assumption that the true WL emission enhancement signals
produced by the WLF will not appear out to the flare ribbon region. It means that only the qualified pixels within the blue contour
identified by each method can be treated as the true WLF-related signals while those outside the blue contour are background WL
emission enhancement signals generated by continuous convective motion. In the end, we can calculate the following two indicators
for quantitative comparison between different methods:
\begin{enumerate}
\item The ratio of qualified pixels within the blue contour to those within the larger red rectangle is denoted as \emph{Indicator a}.
\item The ratio of qualified pixels within the blue contour to all the pixels within the blue contour is calculated as \emph{Indicator b}.
\end{enumerate}

The higher the \emph{Indicator a} a method obtains, the higher the recognition rate of true WLF-related signals among all the
detected signals including background WL emission enhancement ones. \emph{Indicator b} reflects the proportion of WLF-related signals
identified by a method among the potential true ones. Figures \ref{fig4}(a1) and \ref{fig4}(b1) show the variation of \emph{Indicator a}
obtained by five methods under different thresholds for a X2.2-class WLF and a C5.5-class WLF, respectively. It is obvious that
\emph{Indicator a} of Method 3 and Method 5 are generally higher than those of other methods. For a quantitative assessment of
\emph{Indicator a} for the five different methods, we further conducted a statistical analysis on \emph{Indicator a} for 10 WLFs at 6 different
thresholds as shown in panels (a1)-(b2). Every time a method obtains the largest \emph{Indicator a} among the five methods, it will be
marked with score of 5 while the smallest \emph{Indicator a} corresponding to score of 1. Then in Figure \ref{fig4}(c), we plot score
histograms of the five different methods. Same as shown in Figures \ref{fig4}(a1) and \ref{fig4}(b1), Method 3 and Method 5 perform the
best. However, as shown in Figures \ref{fig4}(a2) and \ref{fig4}(b2), \emph{Indicator b} of Method 3 always remains relatively low, which
means the constraint of Method 3 is overly stringent, leading to removal of considerable true WL emission enhancement signals
produced by WLFs. Taking these two indicators both into consideration, we ultimately conclude that Method 5 performs the best among
all the five methods and decide to employ it in the subsequent analysis of WLFs.

To further intuitively compare the effectiveness of the traditional Method 1 and Method 5, as representative of the optimized
methods, we applied the two methods to three flares spanning various GOES energy levels: one X1.6-class WLF (2014-09-10T17:45), one
M1.5-class WLF (2011-12-31T16:26), and one C5.6-class WLF (2015-11-01T03:39). Two aspects will be considered here: the exact identification
of true WL emission enhancement signals produced by WLFs and the optimization of calculated WL emission light curve. It is worth noting
that the M1.5-class flare was classified as non-WLF by \citet{2017ApJ...850..204W}, while it was identified as WLF
by \citet{2020ApJ...904...96C}. As mentioned in the Section \hyperref[sect32]{3.2}, a concept similar to the intrinsic threshold was
also once employed by \citet{2020ApJ...904...96C}, thereby enabling the identification of some WLFs that were unrecognized by the
traditional methods in previous studies.

As shown in Figures \ref{fig5}(a1) and \ref{fig5}(b1), Method 5 identifies a large amount of WL emission enhancement signals
within the X1.6 flare ribbon region covering the main sunspot while Method 1 identifies much fewer signals. Similar results are also
found in Figures \ref{fig6} and \ref{fig7} during the M1.5 and C5.6 flares, indicating the efficiency of Method 5 in capturing WL emission
enhancement signals produced by WLFs. On the other hand, we calculated WL emission light curves of these flares through the two methods as
follows: after determining the qualified pixels in each method, we summed their intensity together for each moment during the flare and
then obtained the WL light curve. As shown in Figures \ref{fig5}(a2) and \ref{fig5}(b2), the WL light curve obtained by Method 5 clearly
shows a significant rapid rise phase and a gradual decay phase while that of Method 1 shows abnormal fluctuations, without a typical
two-phase characteristic. In Figures \ref{fig6} and \ref{fig7}, similar results are also presented. We know that WL light curves will be
significantly influenced by the following two factors: 1) presence or absence of the background WL emission enhancement signals; 2)
sampling completeness of the true WL emission enhancement signals produced by the WLF. As a result, we can conclude that compared with
Method 1, Method 5 can preserve a large amount of faint WLF-related signals and efficiently remove a significant portion of background
WL enhancement signals through the temporal and spatial constraints and the improvement of the threshold.

\section{APPLICATION OF THE OPTIMIZED WLF IDENTIFICATION METHOD}\label{sect4}
According to the results shown in Section \hyperref[sect3]{3}, we prefer to construct a solar WLF sample by applying Method 5
as a representative of the optimized methods proposed here to a flare catalog consisting of 90 flares. This flare catalog comprises 30
C-class, 30 M-class, and 30 X-class flares, among which 24 X-, 30 M-, and 30 C-class ones were randomly chosen from the flare catalog
spanning from 2011 to 2017 as documented in \citet{2021ApJ...917L..29L}. Moreover, we supplemented 6 additional X-class flares from the
period of 2021 to 2023. After corrected for limb darkening and solar rotation, the SDO/HMI continuum images of the 90 solar flares were
analyzed by the Method 5 to identify WLFs. Furthermore, among the identified WLFs, we computed the energy (\emph{E}) and duration ($\tau$)
of over forty WLFs (as shown in Table \ref{t1}) with typical two-phase characteristic of their WL light curve, followed by conducting
statistical analysis.

\subsection{Identification of WLFs}\label{sect41}
Through Method 5, we can obtain the spatial and temporal distribution of WL emission enhancement signals during a flare. Then
according to the typical distribution characteristics of WLF-induced signals shown in Figure \ref{fig1}, we can determine whether the
flare is a WLF or not. Eventually, we identified a total of 9 C-class WLFs, 18 M-class WLFs, and 28 X-class WLFs among the 90 flares.
As shown in Figure \ref{fig8}(a), the percentages of WLF among the C-class, M-class, and X-class flares are 30\%, 60\%, and 93.3\%,
respectively. It is obvious that the higher the GOES energy level of the flares, the greater the proportion of WLF in these flares,
implying that WLFs are more frequently related to energetic solar eruptions like X-class flares \citep{2003A&A...409.1107M,
2009RAA.....9..127W,2017ApJ...850..204W}. However, we also identified 9 WLFs among 30 C-class flares, resulting in an identification
rate of 30\%, which is the highest identification rate of C-class WLFs to date to our best knowledge \citep{2003A&A...409.1107M,
2009RAA.....9..127W,2015SoPh..290.3151B,2018A&A...613A..69S,2018ApJ...867..159S,2020ApJ...904...96C}. We have substantial grounds to
believe that more weak C-class WLFs will be discovered with the development of solar telescopes, which will have higher sensitivity
and resolution. Similar perspectives have already been proposed by previous studies. For instance, \citet{2006SoPh..234...79H} and
\citet{2008ApJ...688L.119J} discovered a C1.6-class WLF and a C2.0-class WLF, respectively, thereby supporting a controversial
hypothesis that all solar flares are WLFs \citep{1989SoPh..121..261N,2006SoPh..234...79H,2008ApJ...688L.119J}.

Additionally, we also investigated the dependence of WLF proportion on the flare types of eruptive and confined ones. In our flare
sample, 32 out of 51 eruptive flares (62.7\%) and 23 out of 39 confined flares (59.0\%) are identified as WLFs. The percentages of WLFs in
these two types of flares are similar and indicate no apparent bias or preference, which is consistent with the result of \citet{2018ApJ...867..159S}.
However, we noticed that the percentage of C-class flares among the eruptive flares studied in \citet{2018ApJ...867..159S} is much lower
than that among the confined flares. As shown in Figure \ref{fig8}(a), the GOES energy level can significantly affect the proportion of WLF.
As a result, we must exclude the energy factor before studying the dependence of WLF proportion on the eruptive and confined types.
In Figure \ref{fig8}(b), it is displayed that the percentages of WLF among confined C-class, M-class, and X-class flares are 34.8\%, 90.9\%,
and 100\%, respectively, while those in eruptive flares are 14.3\%, 42.1\%, 92\%, respectively. It is revealed that the proportion of WLFs in
confined flares is obviously higher than that in eruptive flares across various GOES energy levels.

\subsection{The Energy and Duration of WLFs}\label{sect42}

\begin{table*}
\caption{List of Flares (First 20 events)\label{t1}}
\begin{center}
\begin{tabular}{c c c c c c c c c}
\hline\hline
\textbf{Date} & \textbf{Location\textsuperscript{a}} & \textbf{GOES} & \textbf{SXR peak} & \textbf{Confined/Eruptive} & \textbf{WLFs} & \textbf{WLF peak} & \textbf{WL Duration\textsuperscript{c}} & \textbf{WL Energy\textsuperscript{d}} \\
\textbf{} & \textbf{} & \textbf{Class} & \textbf{time\textsuperscript{b}} & \textbf{Flares} & \textbf{} & \textbf{time} & \textbf{(min)} & \textbf{(10\textsuperscript{29} erg)} \\
\hline
\emph 2011.02.13 & S20E05 & M6.6 & 17:38 & E & yes & 17:35 & 8.78 & 8.73\\
\hline
\emph 2011.02.14 & S20W01 & C9.4 & 12:53 & E & no &  &  & \\
\hline
\emph 2011.02.15 & S20W10 & X2.2 & 01:56 & E & yes & 01:54 & 12.63 & 32.32\\
\hline
\emph 2011.02.16 & S20W28 & C9.9 & 09:11 & C & no &  &  &\\
\hline
\emph 2011.02.19 & N18W11 & C8.5 & 08:04 & C & yes &  &  &\\
\hline
\emph 2011.03.09 & N07W04 & C9.4 & 22:12 & E & no &  &  &\\
\hline
\emph 2011.03.09 & N08W11 & X1.5 & 23:23 & C & yes & 23:21 & 6.60 & 9.10\\
\hline
\emph 2011.08.03 & N14W36 & C8.5 & 19:30 & C & no &  &  &\\
\hline
\emph 2011.08.04 & N16W38 & M9.3 & 03:57 & E & yes & 03:54 & 5.62 & 4.44\\
\hline
\emph 2011.08.09 & N14W69 & X6.9 & 08:05 & E & yes & 08:04 & 14.25 & 132.93\\
\hline
\emph 2011.09.06 & N14W18 & X2.1 & 22:20 & E & yes & 22:19 & 14.90 & 32.19\\
\hline
\emph 2011.09.07 & N14W31 & X1.8 & 22:38 & E & yes & 22:37 & 6.50 & 43.08\\
\hline
\emph 2011.09.08 & N14W41 & M6.7 & 15:46 & C & yes & 15:42 & 12.15 & 8.24\\
\hline
\emph 2011.09.24 & N13E61 & X1.9 & 09:40 & E & yes & 09:36 & 13.23 & 61.93\\
\hline
\emph 2011.10.02 & N10W14 & M3.9 & 00:50 & E & no &  &  &\\
\hline
\emph 2011.11.03 & N21E64 & X1.9 & 20:27 & C & yes &  &  &\\
\hline
\emph 2011.11.06 & N20E29 & C8.8 & 09:56 & C & no &  &  &\\
\hline
\emph 2011.12.05 & S20W05 & C6.9 & 23:25 & C & yes &  &  &\\
\hline
\emph 2011.12.25 & S22W24 & C7.7 & 20:29 & E & no &  &  &\\
\hline
\emph 2011.12.31 & S25E42 & M1.5 & 16:26 & C & yes &  &  &\\
\hline
\end{tabular}
\end{center}
\par
\textbf{Notes.} \\
\textsuperscript{a} The flare location is derived from the \emph{GOES} flare catalog. \\
\textsuperscript{b} Flare peak time of the \emph{GOES} SXR flux. \\
\textsuperscript{c} The decay time of WLFs. \\
\textsuperscript{d} Flare energy radiated in the white light.\\
(This table is available in its entirety in machine-readable form.)
\end{table*}

\begin{figure*}[htbp]
\centering
\includegraphics [width=0.7\textwidth]{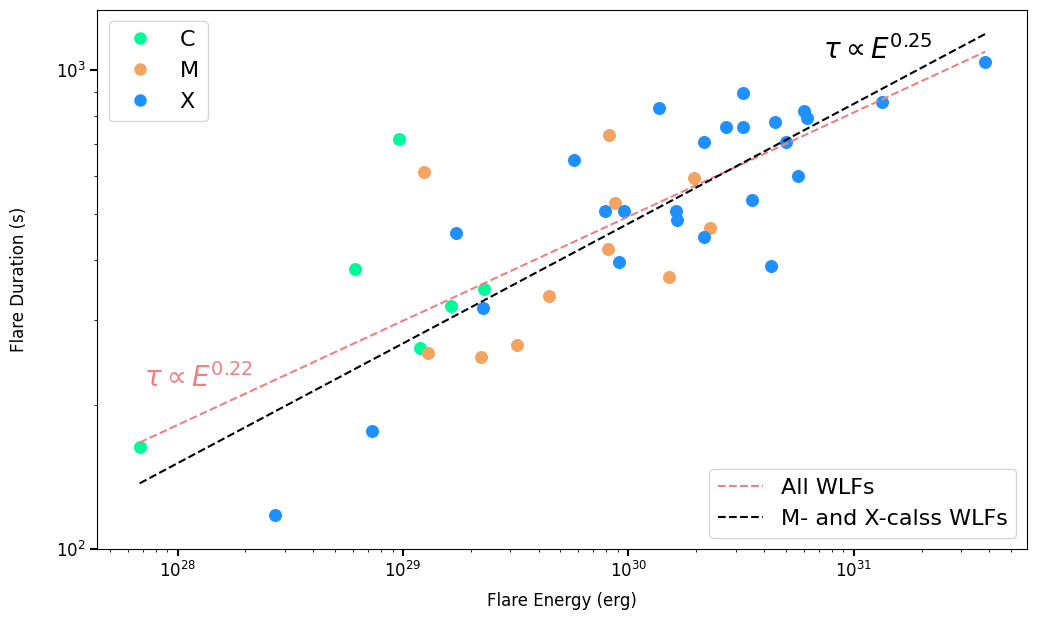}
\caption{Comparison between WLF energy and duration. Points of different colors represent WLFs with different GOES energy
levels, while the red and black dashed lines respectively show the fitting results for all WLFs and for M-class and X-class WLFs using a
linear regression method.}
\label{fig9}
\end{figure*}

As shown in Section \hyperref[sect3]{3}, the optimized method can also obtain more accurate WL light curve of WLFs. Therefore, here
we also calculated energy ($E$) and duration ($\tau$) of the WLFs identified by the Method 5 based on their WL light curves. Before
calculating the energy and duration, we need to preprocess their WL light curves by subtracting the global trend. Based on the original
light curve ($L_{\text{original}}$), we firstly identified the start and end times of each WLF. Then, a linear interpolation was utilized to
replace the segment of the original light curve between the WL start and end times. Subsequently, we applied a smoothing process
(over five data points) to the entire reconstructed light curve for a baseline ($L_{\text{base}}$). Finally, we subtracted these baselines
($L_{\text{base}}$) from the original light curves ($L_{\text{original}}$), resulting in the eventual light curves of WLFs
($L_{\text{wlf}} = L_{\text{original}} - L_{\text{base}}$).

We subsequently employed the same method as in \citet{2013ApJS..209....5S} and \citet{2017ApJ...851...91N} to calculate the energy
($E$) of a solar WLF:
\begin{equation}
\ E=\sigma_{\mathrm{SB}}T_{\mathrm{flare}}^4\int A_{\mathrm{flare}}(t)dt, \label{eq:4}
\end{equation}
\begin{equation}
\ A_{\mathrm{flare}}(t)=\frac{L_{\mathrm{flare}}}{L_{\mathrm{Sun}}}\pi R^2\frac{\int R_\lambda B_\lambda(5800\mathrm{K})d\lambda}{\int R_\lambda B_\lambda(T_{\mathrm{flare}})d\lambda}, \label{eq:5}
\end{equation}
where $\sigma_{\mathrm{SB}}$ stands for the Stefan-Boltzmann constant, $T_{\mathrm{flare}} = 10000K$  means we assume that the
solar flares are radiated by a $T_{\mathrm{flare}} = 10000K$ blackbody \citep{2015SoPh..290.3663K},
$L_{\mathrm{flare}}$/$L_{\mathrm{Sun}}$ is the flare luminosity to the overall solar luminosity, $R$ is the solar radius, $B_\lambda(T)$
is the $Planck$ function at a given wavelength $\lambda$, $R_\lambda$ is a response function of $SDO$/HMI. Furthermore, the
durations ($\tau$) of each flare are calculated as the decay time based on WL peak and WL end times \citep{2015EP&S...67...59M}. It is
noted that before calculating the decay time, we used cubic spline interpolation to interpolate the light curve to one-second intervals,
in order to avoid overestimating the decay time \citep{2017ApJ...851...91N}.

In Figure \ref{fig9}, we used green, orange, and blue points to represent C-class, M-class, and X-class WLFs, respectively.
It is obvious that the energy and duration of these WLFs exhibit a very apparent linear relationship in logarithmic space. A power-law
relation of $\tau \propto {E}^{0.22}$ is obtained by fitting all the WLF data with a linear regression method. It changes to
$\tau \propto {E}^{0.25}$ when only the M-class and X-class WLFs are considered. We find that these two power-law relations are
compatible with the result ($\tau \propto {E}^{0.2-0.33}$) of solar flares observed with HXRs/SXRs \citep{2002A&A...382.1070V,
2008ApJ...677.1385C}, which can be explained by magnetic reconnection theory \citep{2015EP&S...67...59M,2017ApJ...851...91N}.
It is widely accepted that flares are phenomena releasing stored magnetic energies ($E_{mag}$) through magnetic reconnection,
and the duration of flares ($\tau$) is believed to be in accordance with the reconnection time scale ($\tau_{rec}$):
\begin{equation}
\ E\sim fE_{\mathrm{mag}}\sim fB^2L^3, \label{eq:6}
\end{equation}
\begin{equation}
\ \tau\sim\tau_\mathrm{rec}\sim\tau_A/M_A\propto L/\nu_A/M_A, \label{eq:7}
\end{equation}
where $f$ is a fraction of magnetic energy released by a flare, $B$ and $L$ respectively stand for magnetic field strength and length
scale, $\tau_A$ and $\nu_A$ mean the Alfv{\'e}n time and the Alfv{\'e}n velocity , and $M_A$ is the
dimensionless reconnection rate. It is worth noting that the Alfv{\'e}n velocity can be expressed as a function of magnetic field
strength ($B$) and plasma density ($\rho$):
\begin{equation}
\ \nu_{A}=B/\sqrt{4\pi\rho}. \label{eq:8}
\end{equation}
Hence, considering the relationship between $B$, $\rho$ and $\nu_A$, the scaling law about $\tau$ can be derived as follows:
\begin{equation}
\ \tau\propto E^{1/3}B^{-5/3}\rho^{1/2}. \label{eq:9}
\end{equation}
If we consider $B$ and $\rho$ to be relatively consistent across all flares, the relation between $E$ and $\tau$ can be derived as:
 \begin{equation}
\ \tau \propto {E}^{1/3}. \label{eq:10}
\end{equation}

However, it is worth noting that the $E-\tau$ power-law indexes obtained by us slightly deviate from the theoretically ideal
index value of 1/3. We propose the following two interpretations for this discrepancy:
\begin{enumerate}
\item The local properties of flare region, such as $B$ and $\rho$, vary between different flares (or flaring stars).
For instance, \citet{2021ApJ...922L..23A} discovered a power-law relation ($\tau \propto {E}^{0.86 \pm 0.03}$) between the energies and
durations of flares in Kepler-411. The significant deviation of the observed power-law index from the theoretically ideal value of 0.33
indicates that the $E-\tau$ power-law index can be substantially influenced by various local properties of the flare region.
\item A credible $E-\tau$ power-law relation should be derived from a large sample of WLFs, which however is not available here. As a result,
in future study, it is urgent to construct a large sample of solar WLFs and then obtain their accurate parameters like energy and duration.
\end{enumerate}

\section{SUMMARY}\label{sect5}
Based on the continuum intensity data from the \emph{SDO}/HMI, we investigated the typical temporal and spatial characteristics of
WLF-related signals and explored the difference between the background WL emission fluctuation of the sunspot region and that of the
quiet Sun region. According to the related results, we then proposed four optimized WLF identification methods and compared the
effectiveness of these optimized methods as well as the traditional method through a quantitative analysis. Finally, we applied
Method 5, as the representative of optimized methods, to 90 flares with different energy levels to construct a solar WLF sample for
a further statistical study of solar WLF. The main findings are summarized as follows:

\begin{enumerate}
\item WL emission enhancement signals produced by WLFs with different GOES energy levels exhibit clear spatial aggregation feature
and tend to occur repeatedly or consecutively during the flare (Figure \ref{fig1}).

\item The background WL emission fluctuation in sunspot region with strong magnetic fields is significantly weaker than that of the
quiet Sun region (Figures \ref{fig2} and \ref{fig3}). As a result, we proposed a concept of intrinsic threshold according
to the inherent fluctuation of background WL emission caused by constant convective motion for each pixel.

\item Based on the typical temporal and spatial distribution characteristics of WLF-induced signals and the concept of
intrinsic threshold, four optimized WLF identification methods are proposed. According to a quantitative analysis (Figure \ref{fig4}),
we ultimately concluded that Method 5 performs the best among all the five methods.

\item We identified a total of 9 C-class WLFs, 18 M-class WLFs, and 28 X-class WLFs among 90 flares using Method 5. It is
worth noting that the identification rate of C-class WLFs reaches 30\%, which is the highest to date to our best knowledge. It is obvious that the higher
the GOES energy level of the flares, the greater the proportion of WLF in these flares. Furthermore, it is also found that the
percentage of WLFs is higher in confined flares than that in eruptive flares across various GOES energy levels (Figure \ref{fig8}).

\item The energy and duration of these WLFs exhibit a very apparent linear relationship in logarithmic space. A power-law
relation of $\tau \propto {E}^{0.22}$ is obtained by fitting all the WLF data with a linear regression method. It changes to
$\tau \propto {E}^{0.25}$ when only the M-class and X-class WLFs are considered (Figure \ref{fig9}).
\end{enumerate}

It is worth noting that although the improved WLF identification method performs well when applied to X-class and M-class flares, it
does not efficiently identify many weak C-class WLFs. We speculate that it is probably caused by the fact that WL emission
enhancement signals produced by weak WLFs are too weak and short-lived to be detected by the \emph{SDO} observations with
current spatial and temporal resolutions. However, with the launch of the \emph{Chinese H$\alpha$ Solar Explorer}
\citep[\emph{CHASE};][]{2022SCPMA..6589602L} and \emph{Advanced Space-based Solar Observatory} \citep[\emph{ASO-S};][]
{2023SoPh..298...68G}, new solar observations at visible wavelength with higher temporal resolution as well as new channels could be
used for further investigation of solar WLFs. For instance, \citet{2023ApJ...952L...6S} and \citet{2023ApJ...954....7L} investigated
heating mechanism and explosive chromospheric evaporation of solar WLFs by analyzing spectral observations from \emph{CHASE}.
\citet{2024ApJ...963L...3L} investigated flare energy deposition models by analyzing two X-class flares off the solar limb
observed by the White-light Solar Telescope (WST) on the \emph{ASO-S}. \citet{2024SoPh..299...11J} analyzed 205 flares above M1.0 and
identified 49 WLFs at 360 nm from the \emph{ASO-S}/WST observations. Based on these new solar flare observations and optimized WLF
identification methods proposed here, we could have a chance to analyze the response of solar WLFs at different wavelengths, identify
more solar WLFs, and eventually establish a large database of solar WLFs spanning across the X-, M-, and C-classes, which will lay a
base for future statistical studies on solar and stellar WLFs.

For the comparison study of solar and stellar flares, \citet{2017ApJ...851...91N} conducted a statistical study based on 50
solar WLFs and found that the $E-\tau$ power-law relation of solar WLFs ($\tau \propto {E}^{0.38}$) is similar to that of stellar
flares, which can be explained by that they share the same mechanism of magnetic reconnection. However, the observed durations
of stellar flares are found to be an order of magnitude shorter than those predicted from solar WLF observations, which might be
caused by the stronger magnetic field of other stars. It is impressive that the authors then tried to diagnose the magnetic field
strength of flare core region through the observed $E$-$\tau$ relation of flares. But as they mentioned, such attempt is challenging
due to the difficulty in accurate measurements of solar WLFs' emission and the resultant light curve. As a result, in our future
research, we will establish a sufficiently large sample of solar WLFs and attempt to achieve more accurate measurements of solar WLFs'
parameters, which will contribute to address critical questions like: Is WL emission enhancement a common feature of all solar flares?
Can the observed flare $E-\tau$ power-law relation applied as a new probe to local magnetic field of flare source region?

\section*{Acknowledgments}
The authors appreciate the anonymous referee for the valuable suggestions. The data used here are courtesy of the \emph{SDO},
\emph{GOES} and \emph{SOHO} science teams. We appreciate Prof. Ying Li, Dr. Yongliang Song, Yining Zhang, Zixi Guo and
Houle Huang for their helpful suggestions. The authors are supported by the Strategic Priority Research Program of CAS (XDB0560000
and XDB41000000), the National Natural Science Foundation of China (11988101, 11933004, 12273060, and 12222306), the National
Key R\&D Program of China (2022YFF0503800, 2019YFA0405000), the Youth Innovation Promotion Association CAS (2023063), the
Open Research Program of Yunnan Key Laboratory of Solar Physics and Space Science (YNSPCC202211). JFL also acknowledges
support from the New Cornerstone Science Foundation through the New Cornerstone Investigator Program and the XPLORER PRIZE.

\bibliography{ref}{}
\bibliographystyle{aasjournal}

\end{document}